\documentclass[prb,aps,preprint,showpacs]{revtex4}
\usepackage{graphicx}

\begin{document}

\title{Geometrical, electronic and magnetic properties of Na$_{0.5}$CoO$_2$ from first principles}

\author{Zhenyu Li}
\author{Jinlong Yang} \thanks{Corresponding author. E-mail: jlyang@ustc.edu.cn}
\author{J.G. Hou}
\author{Qingshi Zhu}

\affiliation{Hefei National Laboratory for Physical Sciences at
Microscale, Laboratory of Bond Selective Chemistry, and Structure
Research Laboratory, University of Science and Technology of
China, Hefei, Anhui 230026, P.R. China}

\date{\today}

\begin{abstract}

We report a first-principles projector augmented wave (PAW) study
on Na$_{0.5}$CoO$_2$. With the sodium ion ordered insulating phase
being identified in experiments, pure density functional
calculations fail to predict an insulating ground state, which
indicates that Na ordering alone can not produce accompanying Co
charge ordering, if additional correlation is not properly
considered. At this level of theory, the most stable phase
presents ferromagnetic ordering within the CoO$_2$ layer and
antiferromagnetic coupling between these layers. When the on-site
Coulomb interaction for Co $3d$ orbitals is included by an
additional Hubbard parameter $U$, charge ordered insulating ground
state can be obtained. The effect of on-site interaction magnitude
on electronic structure is studied. At a moderate value of $U$
(4.0 eV for example), the ground state is antiferromagnetic, with
a Co$^{4+}$ magnetic moment about 1.0 $\mu_B$ and a magnetic
energy of 0.12 eV/Co. The rehybridization process is also studied
in the DFT+U point of view.

\end{abstract}


\maketitle

\section{INTRODUCTION}

Na$_x$CoO$_2$ was originally studied as a kind of rechargeable
battery material\cite{fouassier75} and thermoelectric
material\cite{terasaki97,terasaki03,wang03}. More recently, the
discovery of the superconductivity in its hydrated x=0.35
compound\cite{takada03}, where the effect of hydration is found to
be limited to lattice expansion\cite{marianetti03,johannes04},
makes it receive a renewed interest. Na$_x$CoO$_2$ has a crystal
structure consisting of 2D triangular lattice Co sheets,
octahedrally coordinated with O above and below the Co planes, and
layers of Na ions sandwiched between the CoO$_2$ sheets. The phase
diagram of Na$_x$CoO$_2$ has been determined by changing the Na
content $x$ using a series of chemical reactions\cite{foo03}.
Their electronic and magnetic properties are found to be strongly
dependent on the number of charge carriers introduced by
deviations from stoichiometry in the Na sublattice.

Theoretical studies on the electronic structure of Na$_x$CoO$_2$
are very active. The early work by Singh
\cite{singh00,singh03}focused on the magnetic properties. He
predicted a weak instabilities of itinerant ferromagnetic
character in range $x=0.3$ to $x=0.7$, with competing but weaker
itinerant antiferromagnetic solution. He used the virtual crystal
approximation, instead of supercell model, to describe the
partially occupation of Na sites. As a result, the possibility of
sodium ion and Co$^{3+}$/Co$^{4+}$ charge ordering is not allowed.
Using pseudopotential plane wave method, Ni \emph{et
al.}\cite{ni03} have optimized the geometries for Na$_x$CoO$_2$
($x=0.25$, 0.5, 0.75 and $x=1$) with a $2 \times 2 \times 1$
supercell. They found that the itinerant ferromagnetism is
strongly suppressed by the local distortions of the oxygen around
the cobalt. With optimized geometries, they identified a phase
transition from wide-band ferromagnetic to narrow band
paramagnetic metals with $x$ increasing. By carefully examining
the doped electron density, Marianetti \emph{et
al.}\cite{marianetti03} found that the rigid band model is not
suitable for this Na doped system. A rehybridization driven by a
competition between the on-site Coulomb interaction and the
$e_g$-oxygen hybridization is identified. They have also used a
modified Hubbard model to study this kind of rehybridization.
Within the framework of density functional theory (DFT), the
on-site Coulomb correlation effects can be dealt with DFT+U
method\cite{anisimov97}. Using this method, Zou \emph{et
al.}\cite{zou03} have studied the electronic structure of
Na$_x$CoO$_2$ with varying $c/a$ ratio. They did not find critical
change of the band structure near $E_F$, except a notable decrease
of the DOS at $E_F$. To consider the relations between correlation
and charge ordering, Pickett and coauthors\cite{kunes03,lee04}
have applied the DFT+U method to a
$\sqrt{3}\times\sqrt{3}\times\frac{1}{2}$ supercell of
Na$_x$CoO$_2$ at $x=1/3$ and $x=2/3$. They found that the
parameter $U$ is critical to charge and spin orderings (for small
supercell, the so called charge disproportionation always leads to
charge ordering). They suggested that the phase diagram of
Na$_x$CoO$_2$ is characterized by a crossover from effective
single-band character with $U\gg W$ for $x>0.5$ into a three-band
regime for $x<0.5$, where $U\sim W$ and correlation effects are
substantially reduced.

Na$_{0.5}$CoO$_2$ is of special interests among the Na$_x$CoO$_2$
family. In the recent experimental phase diagram of Na$_x$CoO$_2$,
a novel sodium ion ordered insulating state at $x=0.5$ was
found\cite{foo03}, with the most strongly developed
superstructure\cite{huang04,zandbergen04}. This insulating state
separates two kinds of metallic phase, \emph{i.e.} Curie-Weiss
metal for $x>0.5$ and paramagnetic metal for $x<0.5$. It is
suggested that the occurence of the insulating state indicates a
strong interaction between the ions and holes even though they
occupy separate layers, and a charge ordering is thought to
accompany the sodium ion ordering\cite{foo03}. Therefore it is
interesting to compare the electronic structure based on this
experimental sodium ion ordered structure with previous
theoretical results to identify the effect of Na ordering. For
example, it will be interesting to see if the Na ordering will
produce a Co charge ordering at DFT level. On the other hand, the
correlation effects of Na$_{0.5}$CoO$_2$ is considered to be a
betweenness comparing to small values for $x>0.5$ and relatively
large values for $x<0.5$\cite{ni03,lee04}, which demands a study
at the DFT+U level to identify the effect of correlation.
Unfortunately, there is no theoretical work in the literature for
this important $x=0.5$ case with both Na ordering and correlation
effects being properly considered. In this paper, we report the
electronic structure of Na$_{0.5}$CoO$_2$ focusing on the
possibility of charge ordering and based on the experimental
superstructure at both DFT and DFT+U level.

\section{COMPUTATIONAL METHOD}

The calculations in this work were performed with the Vienna
\emph{Ab initio} Simulation Package
(VASP)\cite{kresse96,kresse96b}, which is a first-principles
plane-wave code, treating the exchange and correlation in the DFT
scheme. In this study, the Perdew-Wang functional
form\cite{perdew92} of generalized gradient approximation (GGA)
was used. For spin-polarized calculations, the spin interpolation
of Vosko \emph{et al.}\cite{vosko80} was also adopted. The
projector augmented wave (PAW)\cite{blochl9453} method in its
implementation of Kresse and Joubert\cite{kresse99} was used to
describe the electron-ion interaction. The Kohn-Sham equations
were solved via a Davidson-block iteration scheme\cite{kresse96}.

A \(\sqrt{3}\times2\times1\) supercell was introduced to describe
the experimental sodium ion ordering. Brillouin zone (BZ)
integrations were performed on a well converged grid of \( 8\times
8\times 4 \) Monkhorst-Pack \cite{monkhorst76} special points. The
total energy and density of states (DOS) were calculated using the
linear tetrahedron method with Blochl
corrections\cite{blochl9423}. The plane wave kinetic energy cutoff
was fixed to 500 eV. Dudarev \emph{et al.} 's
approach\cite{dudarev98} was adopted for DFT+U calculations. Since
this DFT+U functional depends only on the difference of Hubbard
parameter $U$ and screened exchange parameter $J$, $J$ was fixed
to 1 eV during all DFT+U calculations. The Hubbard on-site Coulomb
interaction is applied to Co $3d$ orbitals only. Geometry
relaxation is performed at a spin unpolarized GGA level.

\section{RESULTS AND DISCUSSION}

\subsection{Geometrical properties}

A \(\sqrt{3}\times2\times1\) superstructure of Na$_{0.5}$CoO$_2$
has been found by powder neutron diffraction
experiment\cite{huang04}. This superstructure has the $Pnmm$ space
group symmetry, with a cell formula: Na$_4$Co$_8$O$_{16}$. In this
experimental structure model, the Na1 and Na2 sites ($2b$ and $2d$
Wyckoff sites for original $P6_3/mmc$ space group, and $2b$ and
$2a$ sites for this $Pnmm$ space group respectively) are equally
occupied, and the ordered Na ions form one-dimensional zigzag
chains. Two types of Co ions ($4f$ and $4d$), which differ subtly
in their coordination by oxygen, are also in chains. The existence
of two kinds of Co sites allows the emergence of two inequivalent
Co ions, namely Co$^{3+}$ and Co$^{4+}$, in the process of
self-consistency. Therefore this structure model can be used to
study the possibility of charge ordering.

The geometry optimization, starting from the experimental
structure, reaches its convergence until all forces vanished
within 0.01 eV/\AA. The optimized geometry is shown in Fig.
\ref{geometry}. During the optimization, the cell parameters are
fixed to the experimental values ($a=4.876$\,\AA, b=5.630\,\AA,
and $c=11.063$\,\AA), which results in only a neglectable stress
less than 20 kB. As listed in Table \ref{tbl:geo}, relaxations of
atomic positions from the experimental ones are also found to be
very small. The largest relaxation comes from the first two Na
atoms along $x$ direction (0.06 \AA), typical relaxations are
about 0.01$\sim$0.02\,\AA. The calculated two kinds of Co-O
distances (1.91 \AA\, and 1.88 \AA) are very similar to
experimental values (1.90 \AA\, and 1.86 \AA). In an ideal
structure model of Na$_x$CoO$_2$, Co and O may form layers of
edge-shared CoO$_6$ octahedra in a triangular lattice, but
distortion of the octahedra with the variation of the O hight
$z_0$ from its ideal value is always exists in real system. Our
optimized $z_0$ varies from 0.0845 $c$ to 0.0892 $c$ for different
O sites, resulting in the Co-O-Co angles varying from 96.2$^\circ$
to 97.0$^\circ$. Comparing to the 90$^\circ$ angle in the
undistorted octahedra, the optimized structure shows a
considerable distortion of the oxygen octahedra. For
\(2\times2\times1\) supercell, we get a slightly smaller
distortion, but Ni \emph{et al.}\cite{ni03} get a significantly
larger one. We notice that they have used a relatively small
kinetic energy cutoff in their study.

\subsection{GGA electronic structure}

We can clearly identify three well separated manifolds in the
paramagnetic GGA band structure and DOS of Na$_{0.5}$CoO$_2$, as
shown in Fig. \ref{ggaband}. The lowest in energy is the O $2p$
bands, with a bandwidth about 5.5 eV (only part of them are shown
in the figure). The other two manifolds come from the Co $3d$
orbitals, which are splitted into a lower lying $t_{2g}$ manifold
and an upper lying $e_g$ manifold, separated by approximately 2.5
eV according to the oxygen octahedral crystal field. Although
these three manifolds are well separated in energy, the
hybridizations of O $2p$ states with both the $t_{2g}$ and $e_g$
states are notable. The Fermi energy locates at the upper edge of
the $t_{2g}$ manifold, indicating a metallic GGA ground state.

The low energy properties of Na$_{0.5}$CoO$_2$ will be determined
by the $t_{2g}$ manifold, which has a bandwidth of 1.6 eV. The
trigonal symmetry of the Co sites further splits the $t_{2g}$
states into one with $a_g$ symmetry and a doubly degenerate
$e_g^\prime$ pair. In Fig. \ref{ggaband}, we also plot the $a_{g}$
(Co $d_{z^2}$) partial DOS. As found by Lee \emph{et
al.}\cite{lee04}, the $a_{g}$ states have a bandwidth almost
identical to that of the whole $t_{2g}$ manifold, which indicates
that the widely used single-band model may not be sufficient
enough to describe the GGA electronic structure of this material.

Fermi surface often plays an important role in the electronic
structure of materials. Our GGA Fermi surfaces for the five bands
across Fermi level are shown in Fig. \ref{fs}, which are obtained
by a \(32\times28\times14\) BZ sampling and viewed by XCrySDen
package\cite{kokalj99}. LDA/GGA band structure is expected to give
valuable and accurate information about the Fermi surface even in
system with strong correlations\cite{johannes04}. In our supercell
structure, there will be band folding from the first BZ of the
original hexagonal lattice (black hexagon in the Fig. \ref{fs}a),
and the Fermi surface thus becomes very complex. We notice that,
on the other hand, the band structure of the superstructure
geometry is far from simply band folding, for example, the Fermi
surface of the fourth band (band 99) shows a prominent 3D
character, which has not been reported in previous virtual crystal
study\cite{singh00,singh03}. According to the complexity of the
present Fermi surfaces, nesting effect for Na$_{0.5}$CoO$_2$ may
not be very strong, and we thus don't expect to find corresponding
spin density wave (SDW) state.

Spin-polarized calculations are carried out in order to address
the possibility of magnetic ordering. When the ordering is
constrained to be ferromagnetic (FM), a stable low spin solution
is obtained, with magnetic moments of 0.4 and 0.45 $\mu_B$ for Co1
and Co2 respectively. But the corresponding energy gain from
ferromagnetic instability is only 26 meV/Co. The moment values are
somewhat smaller than that (0.5 $\mu_B$) reported by
Singh\cite{singh00} under virtual crystal approximation, which may
due to the difference between their structural model and ours. In
our structure, the triangular symmetry is slightly broken.

Antiferromagnetic (AFM) ordering is frustrated on the 2D
triangular lattice. But if the magnetic moments on a sublattice
vanish, the left other sublattice may geometrically permit the
presence of AFM ordering. Here we get an AFM model (AFM1) by
fixing the magnetic moments at Co1 sites to be zero. As shown in
Fig. \ref{geometry}, the Co2 sites form a 1D chain in the $x$-$y$
plane, and it is easy to establish AFM ordering within this chain.
In a unit cell, the two AFM chains are set to be out of phase. The
AFM1 model results in a magnetic moments of Co2 ions about 0.39
$\mu_B$, which is much larger the result (0.21 $\mu_B$) for the
Singh's AFM model\cite{singh00}. We notice that in that model each
Co ion has also two nearest neighbors of like spin in addition to
four of opposite spin. Another AFM model (AFM2) is constructed by
alter the direction of magnetic moments of neighboring FM Co
layers. Although the interlayer exchange interaction is expected
to be small, we get lower energy for AFM2 state than that of FM
state.

Until now, we get a metallic ground state with intra-layer FM
ordering and inter-layer AFM ordering. Comparing with the results
of Singh\cite{singh00}, we may conclude that Na ordering alone
will not qualitatively affect the GGA electronic structure to
generate the band gap and charge ordering observed in experiments.
For the AFM1 state, where charge ordering is enforced, the
magnetic energy (7 meV/Co) is much smaller than those of AFM2 and
FM states, which indicates the enforced charge ordering under GGA
level is unfavorable in energy. This result is consistent with the
calculations for Na$_{1/3}$CoO$_2$, where attempts using LDA to
obtain self-consistent AFM spin ordering always converge to the FM
or nonmagnetic solution\cite{kunes03}. This discrepancy between
theory and experimental observation naturally leads to a further
consideration of correlation effect.

\subsection{DFT+U electronic structure}

DFT+U\cite{anisimov97,dudarev98} is a kind of method aiming to
deal with the on-site Coulomb interaction within the framework of
DFT. It treats the local electrons in a Hartree-Fock manner, which
drives local orbital occupations to integral occupancy as $U$
increases. For transition metal oxide, one may find a larger split
of $d$ orbitals for larger $U$, as shown in Fig.
\ref{dosfm}-\ref{dosafm2}, where partial DOS of cobalt $3d$
orbitals and oxygen $2p$ orbitals are plotted for FM, AFM1 and
AFM2 states respectively. The cobalt $3d$ partial DOS for Co1 and
Co2 sites are separated. We do not present the results for spin
restricted paramagnetic states in this section, because the energy
differences between these states and corresponding magnetic states
increase rapidly with $U$ (0.4 eV/Co for $U=7.0$ eV). The
paramagnetic states are still metallic even with $U$ as large as
7.0 eV.

In Fig. \ref{dosfm}, the top panel is the GGA result, and the
three below panels represent three typical electronic structure
behaviors for different $U$ values. For small $U$ (see panel b),
the electronic structure is similar to the above GGA one, except
for a strong trend towards gap opening and charge ordering. As
shown in Fig. \ref{moment}, the moments on the two inequivalent Co
sites are nearly equal and also similar to the GGA values (which
is the $(U-J)\to 0$ limit) until $U=U_c=$3 eV. Above $U_c$, the
unoccupied part of the $a_g$ states weight more and more for Co2
states with the increase of $U$. Disproportionation from
Co$^{3.5+}$ into $s=\frac{1}{2}$ Co$^{4+}$ and $s=0$ Co$^{3+}$
ions is nearly complete at $U=4$ eV and is accompanied by a metal
insulator (Mott-like) transition from conducting to insulating.
This insulating phase results from the splitting of the $a_g$
states, which lead to an unoccupied narrow band containing one
hole for each Co2 ions and an occupied band on the Co1 ion, as
shown in Fig. \ref{dosfm}c. We judge the formation of charge
ordering from local magnetic moments instead of from charge
population directly, because the charge difference between Co1 and
Co2 are relatively small (only 0.1$\sim$0.2 electron). This is a
result of the hybridization of the O $2p$ and Co $3d$ orbitals. In
fact, oxygen may directly contribute to the charge ordering in
some oxides, as suggested by Coey\cite{coey04}. We also notice
that the charge ordering and gap opening with a moderate $U$ is
far from obvious, since Zou \emph{et al.}\cite{zou03} have gotten
a metallic ground state with a relatively large Coulomb $U$ (5 eV)
but without adopting any supercell structure. Therefore the sodium
ion ordering is critical to the insulating ground state.

There is no clear boundary for moderate and large values of $U$,
but a relatively large $U$ value will feature several characters
as shown in Fig. \ref{dosfm}d. First, the empty $a_g$ band are
pushed further higher, which make it strongly mix with the upper
$e_g$ bands. Secondly, the large splitting of the Co $3d$ orbitals
even squeeze some Co2 $3d$ states to the bottom of the O $2p$
manifold, and consequently the O $2p$ character at the edge of
valence band near Fermi energy has exceeded the Co $3d$
characters. Thirdly, the magnetic moments of cobalt are exceed 1
$\mu_B$ in this region (up to 1.42 $\mu_B$ per Co for $U=7.0$ eV),
which seems curious at first sight. We now give two possible
rationalizations for this observation. One interpretation comes
from the population analysis process. We notice that oxygen is
also spin polarized in our calculations for large $U$, which makes
the net magnetic moment of the whole supercell is still about 1.0
$\mu_B$ per Co. Therefore the nominal magnetic moment being larger
than 1 $\mu_B$ per Co may be just an artifact of PAW population
analysis as a result of the strong hybridization between Co and O.
Another possibility is that the large on-site Coulomb interaction
may really promote a transition from low spin state to a
intermediate spin state as demonstrated by Korotin {\it et
al.}\cite{korotin96} for Co ions in vertex sharing CoO$_6$
octahedra. This type of transition is consistent with the strong
mixing of empty $a_g$ and $e_g$ bands.

The evolution of electronic structures with $U$ for AFM states is
generally similar to that for FM state. The most significant
difference is that the charge ordering occurs well before gap
opening for AFM1 state. As shown in Fig. \ref{moment}, the
Co$^{4+}$ moment grows immediately for AFM1 state as $U$ increases
from 1 eV, but the gap opens until $U=3$ eV, which is different
from the results of Lee \emph{et al.}\cite{lee04} for
Na$_x$CoO$_2$ with $x=1/3$ and $x=2/3$.

The energy difference between the FM and AFM phases as a function
of $U$ is plotted in Fig. \ref{moment}c. The general trend gives
lower energy AFM state for small $U$ and lower energy FM state for
large $U$, except the relatively higher AFM1 energy for small $U$,
which is caused by the artificial charge ordering as previously
discussed. For large $U$, the lower energy of FM state is a
natural result of the weakening of superexchange effect, and it is
also consistent with the prominent DOS of the majority spin
component at deep energy as shown in Fig. \ref{dosfm}d. The energy
difference between FM and AFM2 is generally small according to the
relatively small exchange interaction between different layers.
According to our result, the magnitude of on-site Coulomb
interaction $U$ in this material can be probed by experimental
magnetic measurement.

From Fig. \ref{dosfm}-\ref{dosafm2}, we can find an interesting
splitting and ordering of the $e_g$ orbitals. With the increase of
$U$, the lower part of the $e_g$ orbitals becomes mainly from Co1
sites and the upper part from Co2 sites. We notice that this kind
of orbital ordering of unoccupied $e_g$ orbitals will not affect
the low energy properties of Na$_{0.5}$CoO$_2$. Although it has
been found in LaCoO$_3$\cite{korotin96}, orbital ordering of
valence band orbitals is not very possible in this structure,
since they are not doubly degenerated.

A rehybridization process is proposed to describe the competition
between the on-site Coulomb interaction and the $e_g$-oxygen
hybridization in doped CoO layers\cite{marianetti03}. This process
can be seen clearly from the partial DOS of O $2p$ orbitals as
shown in Fig. \ref{dosfm}-\ref{dosafm2}. At GGA level,
hybridization of oxygen $2p$ states and $e_g$ states are more
significant than the $t_{2g}$-oxygen hybridization. With the
increase of $U$, the on-site Coulomb interaction becomes stronger,
which can be minimized by unmixing the oxygen and $e_g$ orbitals
in order to decrease the occupation of the $e_g$ orbitals. In Fig.
\ref{dosfm}-\ref{dosafm2}, we can clearly see this kind of
unmixing by the decrease of oxygen $2p$ partial DOS in the $e_g$
manifold. For large value of $U$, the hybridization of oxygen $2p$
states and $t_{2g}$ states is somewhat enhanced, which is not
included in the modified Hubbard model study by Marianetti
\emph{et al.}\cite{marianetti03}.

We have studied the effect of Hubbard $U$ on the electronic
structures of Na$_{0.5}$CoO$_2$ in the previous part of this
section, but the actual value of $U$ in this material is not
determined yet, of which very different values are proposed in the
literature\cite{singh00,zou03,lee04,chainani03}. Based on the
folowing two experimental observes, we estimate that the vaule of
$U$ should be moderate, as also suggested by Lee \emph{et
al.}\cite{lee04}. First, the insulating ground state suggests that
the value of $U$ should not be too small. Secondly, Na$_x$CoO$_2$
with $x$ away from 0.5 is always metallic indicates a narrow
energy gap for Na$_{0.5}$CoO$_2$, which prefer a not too large
value of $U$. Since there was experimental report\cite{chou03} on
electron mass enhancement for small $x$ of Na$_x$CoO$_2$, people
have compared the experimental electronic specific heat
coefficient $\gamma$ with the band value to estimate the strength
of correlation\cite{lee04,singh00}. But in the Na$_{0.5}$CoO$_2$
case, where Hubbard type correlation drives system to a Mott
insulator, $\gamma$ will reduced by the correlation effects. In
fact, the experimental $\gamma$ is only 3
mJ/mol-K$^2$,\cite{huang04} which is much smaller than our GGA
band value (10.5 mJ/mol-K$^2$), but very near to the DFT+U value
(zero).

At a plausible value of $U$, 4 eV for example, intralayer AFM
coupling is favored over FM and interlayer AFM coupling (ref Fig.
\ref{moment}c). We notice that the magnetic state in Na$_x$CoO$_2$
is far from being clearly determined in experiments. There are
some experiments supporting AFM
coupling\cite{sakurai03,fujimoto03}, while there are also some
recent experiments supporting FM
coupling\cite{sakurai03,boothroyd04,mothashi03}. With this
intermediate value of $U$, the ground state (AFM1) magnetic moment
for Co2 is 1.0 $\mu_B$, and the magnetic energy is 0.123 ev/Co.

\section{CONCLUSION}

Based on DFT+U method, we have studied the effects of Na ordering
and on-site Coulomb correlation on the geometrical, electronic and
magnetic properties of Na$_{0.5}$CoO$_2$. Without including
on-site Coulomb interaction $U$, the experimental insulating phase
can not be produced by DFT. For small value of $U$, ground state
is still metallic. At this time, no charge ordering occur for FM
coupling, and symmetry enforced charge ordering for intralayer AFM
coupling is unfavorable in energy. Increasing parameter $U$ to a
moderate value can open an energy gap and form charge ordering for
both FM and AFM states. Large $U$ value will largely split the Co
$3d$ orbitals, which leads to squeezing Co $3d$ states to the
bottom of O $2p$ manifold, mixing of empty $a_g$ and $e_g$ states,
and large Co magnetic moments.

Accompanying with the early work on Na$_x$CoO$_2$ ($x=1/3$ and
$x=2/3$)\cite{kunes03,lee04}, our study will shed some insights
into understanding of the complex interplay of Na content,
superstructure, correlation effects and tendency of various type
of ordering in Na$_x$CoO$_2$. However, a complete understanding on
this issue, for example, why the $x=2/3$ case with stronger
commensurability effect and maybe also stronger correlation is not
a charge ordered insulator in experiment, will need much further
work.

\begin{acknowledgments}

This work is partially supported by the National Project for the
Development of Key Fundamental Sciences in China (G1999075305,
G2001CB3095), by the National Natural Science Foundation of China
(50121202, 20025309, 10074058), by the Foundation of Ministry of
Education of China, and by ICTS, CAS.

\end{acknowledgments}

\clearpage
\begin{figure}
\caption{(Color online) The optimized crystal structure of
Na$_{0.5}$CoO$_2$. The designations of the Na and Co atoms are as
indicated.} \label{geometry}
\end{figure}

\begin{figure}
\caption{GGA paramagnetic band structure (left panel) and density
of states (DOS)(right panel) of Na$_{0.5}$CoO$_2$. The Fermi
energy is at zero. $\Gamma$=(0,0,0), M=(1/2,0,0), N=(1/2,1/2,0),
and A=(0,0,1/2). The dot line in the DOS panel represents the Co
$a_g$ partial DOS.} \label{ggaband}
\end{figure}

\begin{figure}
\caption{(Color online) (a) Schematic diagram of the reciprocal
lattice and the first Brillouin zone for the original $P6_3/mmc$
lattice (black) and the superstructure adopted in this work (red)
respectively. (b)-(f) Top views of Fermi surface within the first
Brillouin zone for band 96 to band 100 of Na$_{0.5}$CoO$_2$
respectively. } \label{fs}
\end{figure}

\begin{figure}
\caption{(Color online) Density of states (DOS) for FM phase.
Total DOS, O $2p$ partial DOS, and $3d$ partial DOS of Co1 and Co2
for (a) GGA result and $U=$ (b) 3.0, (c) 4.0, and (d) 7.0 eV are
presented.} \label{dosfm}
\end{figure}

\begin{figure}
\caption{(Color online) Density of states (DOS) for AFM1 phase.
Total DOS, O $2p$ partial DOS, and $3d$ partial DOS of Co1 and Co2
for (a) GGA result and $U=$ (b) 3.0, (c) 4.0, and (d) 7.0 eV are
presented.} \label{dosafm1}
\end{figure}

\begin{figure}
\caption{(Color online) Density of states (DOS) for AFM2 phase.
Total DOS, O $2p$ partial DOS, and $3d$ partial DOS of Co1 and Co2
for (a) GGA result and $U=$ (b) 3.0, (c) 4.0, and (d) 7.0 eV are
presented.} \label{dosafm2}
\end{figure}

\begin{figure}
\caption{(Color online) Effect of the intra-atomic repulsion $U$
on (a) magnetic moments, (b) energy gap and (c) energy difference
between AFM and FM states of Na$_{0.5}$CoO$_2$. The energy of FM
state is set to zero.} \label{moment}
\end{figure}

\begin{table}
\caption{Optimized structral parameters. Experimental values are
presented in parentheses.} \label{tbl:geo}
\begin{tabular}{ccrlrlrl}
\hline\hline
Atom & Site & \multicolumn{2}{c}x & \multicolumn{2}{c}y & \multicolumn{2}{c}z \\
\hline
Co1 & 4$f$ & 0.003&(0) & 0.250&(1/4) & 0.003&(0)  \\
Co2 & 4$d$ & 0.500&(1/2) & 0.000&(0) & 0.000&(0)  \\
Na1 & 2$b$ & -0.030&(-0.018) & 0.250&(1/4) & 0.750&(3/4)  \\
Na2 & 2$a$ & 0.355&(0.352) & 0.750&(3/4) & 0.750&(3/4) \\
O1  & 4$f$ & 0.336&(1/3) & 0.250&(1/4) & 0.0880&(0.0895) \\
O2  & 4$f$ & 0.330&(1/3) & 0.750&(3/4) & 0.0845&(0.0820) \\
O3  & 8$g$ & -0.166&(1/6) & 0.000&(0) & 0.0892&(0.0895) \\
\hline\hline
\end{tabular}
\end{table}

\clearpage
\begin{center}
{\Large Figure 1, Li \emph{et al.}}
\end{center}
\vspace{2cm}
\includegraphics[keepaspectratio,totalheight=11cm]{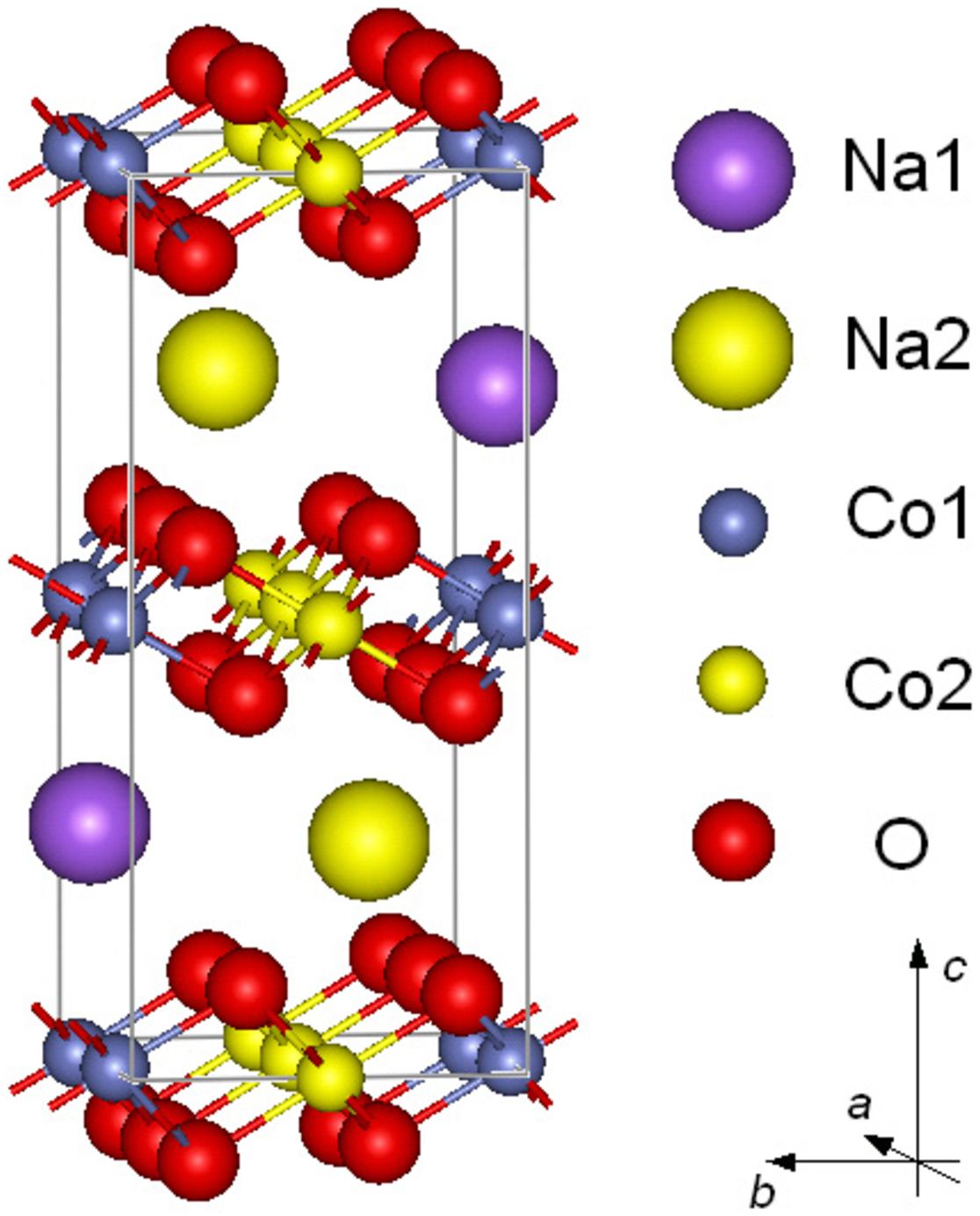}

\clearpage
\begin{center}
{\Large Figure 2, Li \emph{et al.}}
\end{center}
\vspace{2cm}
\includegraphics{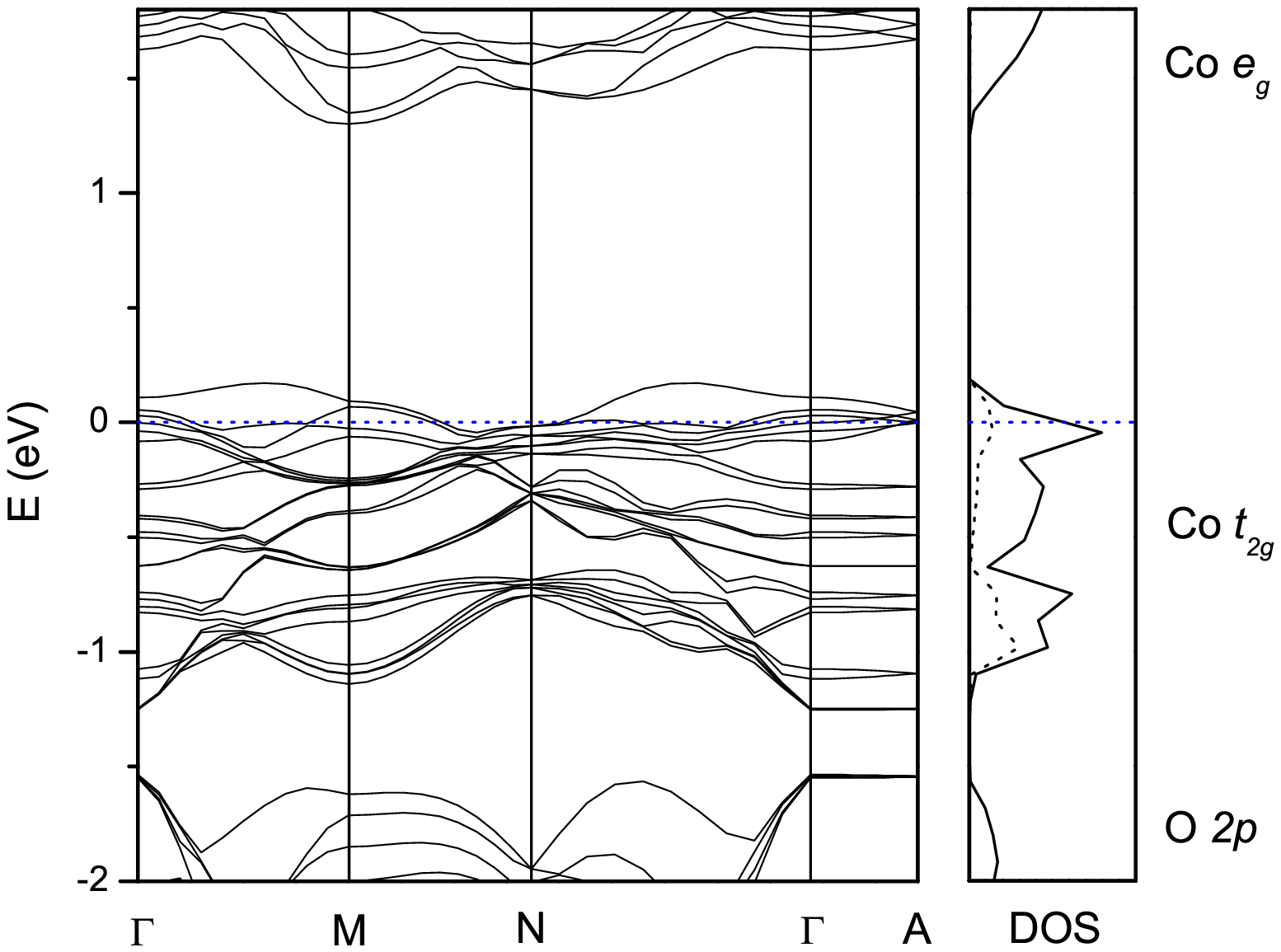}

\clearpage
\begin{center}
{\Large Figure 3, Li \emph{et al.}}
\end{center}
\vspace{2cm}
\includegraphics[keepaspectratio,totalheight=14cm]{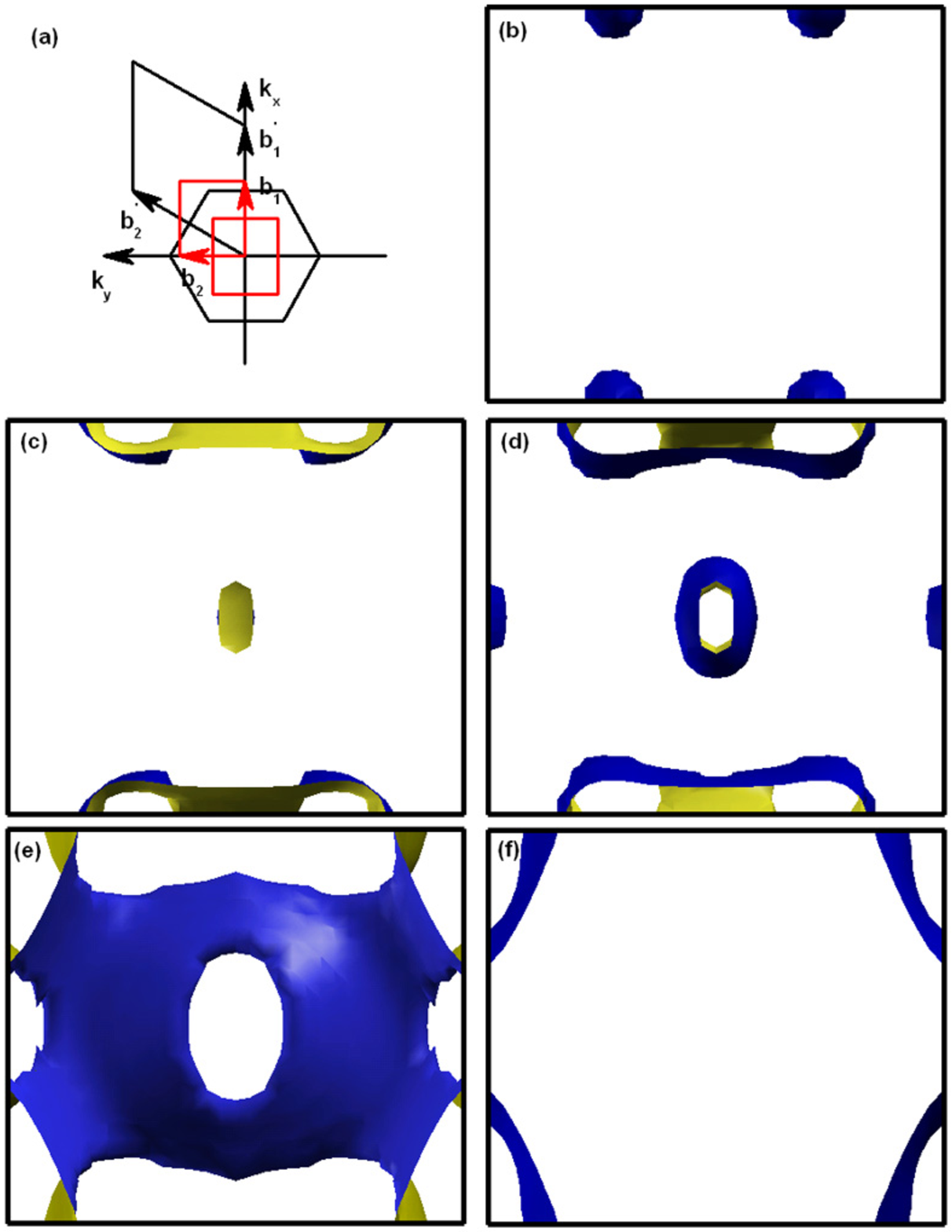}

\clearpage
\begin{center}
{\Large Figure 4, Li \emph{et al.}}
\end{center}
\vspace{2cm}
\includegraphics[keepaspectratio,totalheight=18cm]{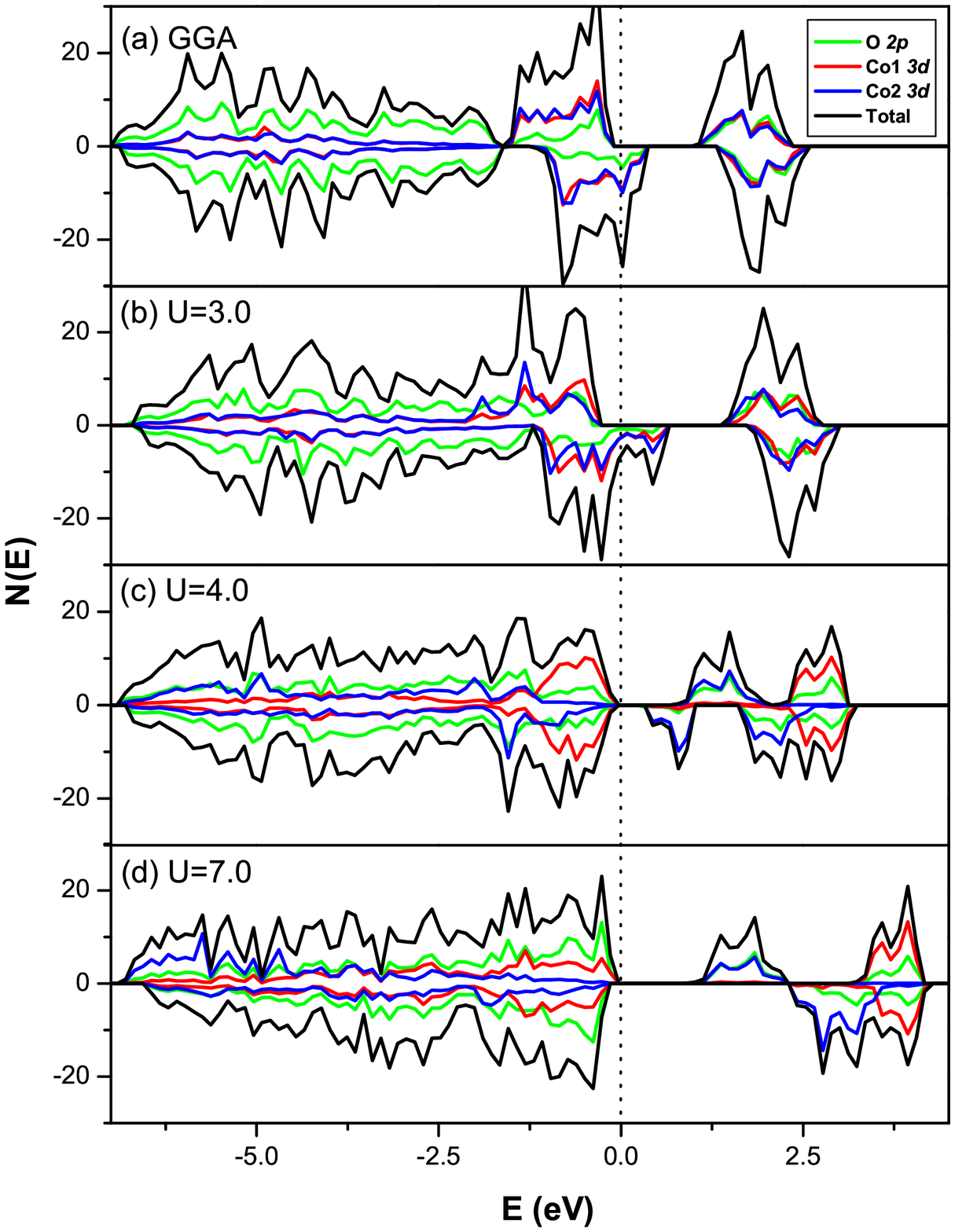}

\clearpage
\begin{center}
{\Large Figure 5, Li \emph{et al.}}
\end{center}
\vspace{2cm}
\includegraphics[keepaspectratio,totalheight=18cm]{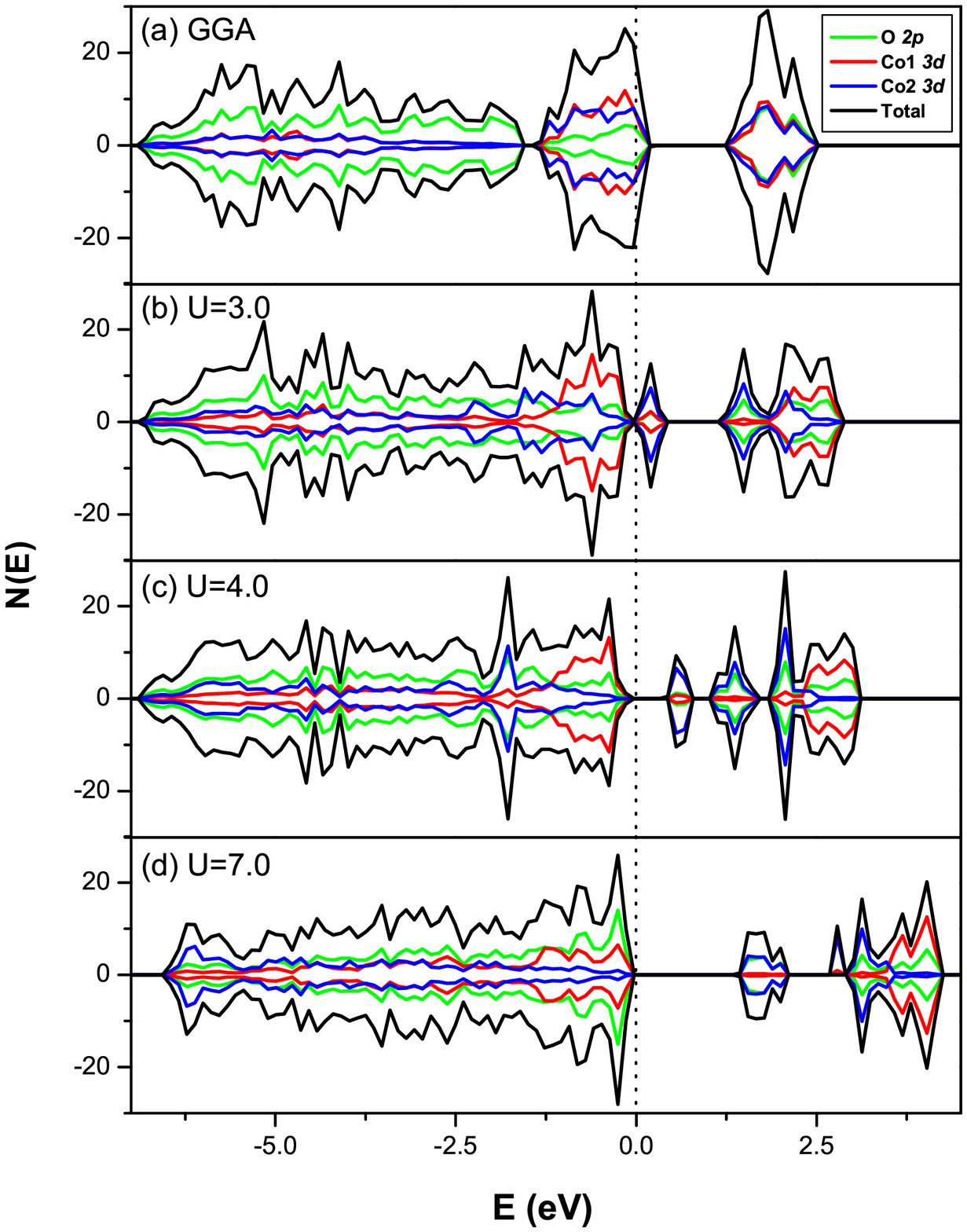}

\clearpage
\begin{center}
{\Large Figure 6, Li \emph{et al.}}
\end{center}
\vspace{2cm}
\includegraphics[keepaspectratio,totalheight=18cm]{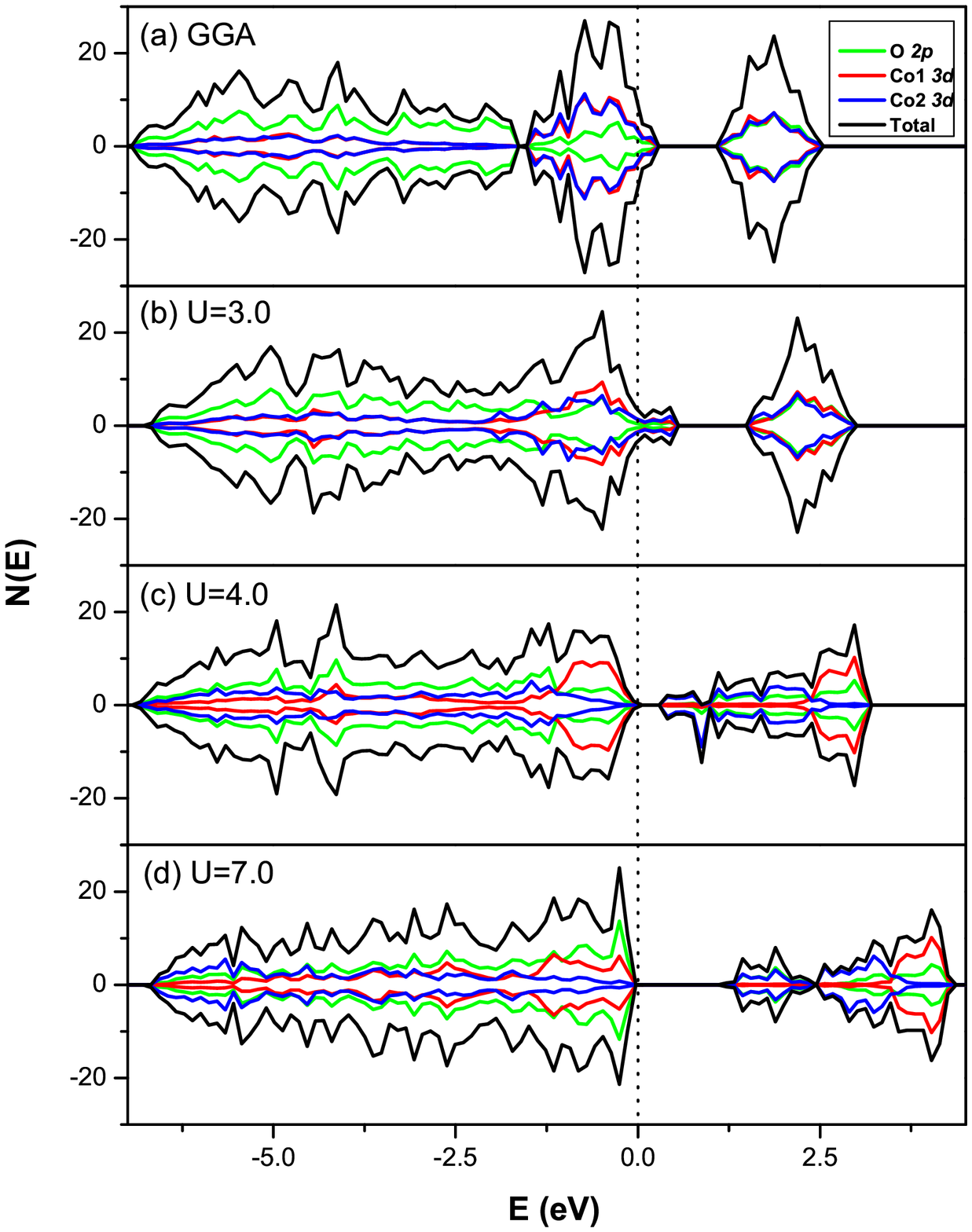}

\clearpage
\begin{center}
{\Large Figure 7, Li \emph{et al.}}
\end{center}
\vspace{2cm}
\includegraphics[keepaspectratio,totalheight=18cm]{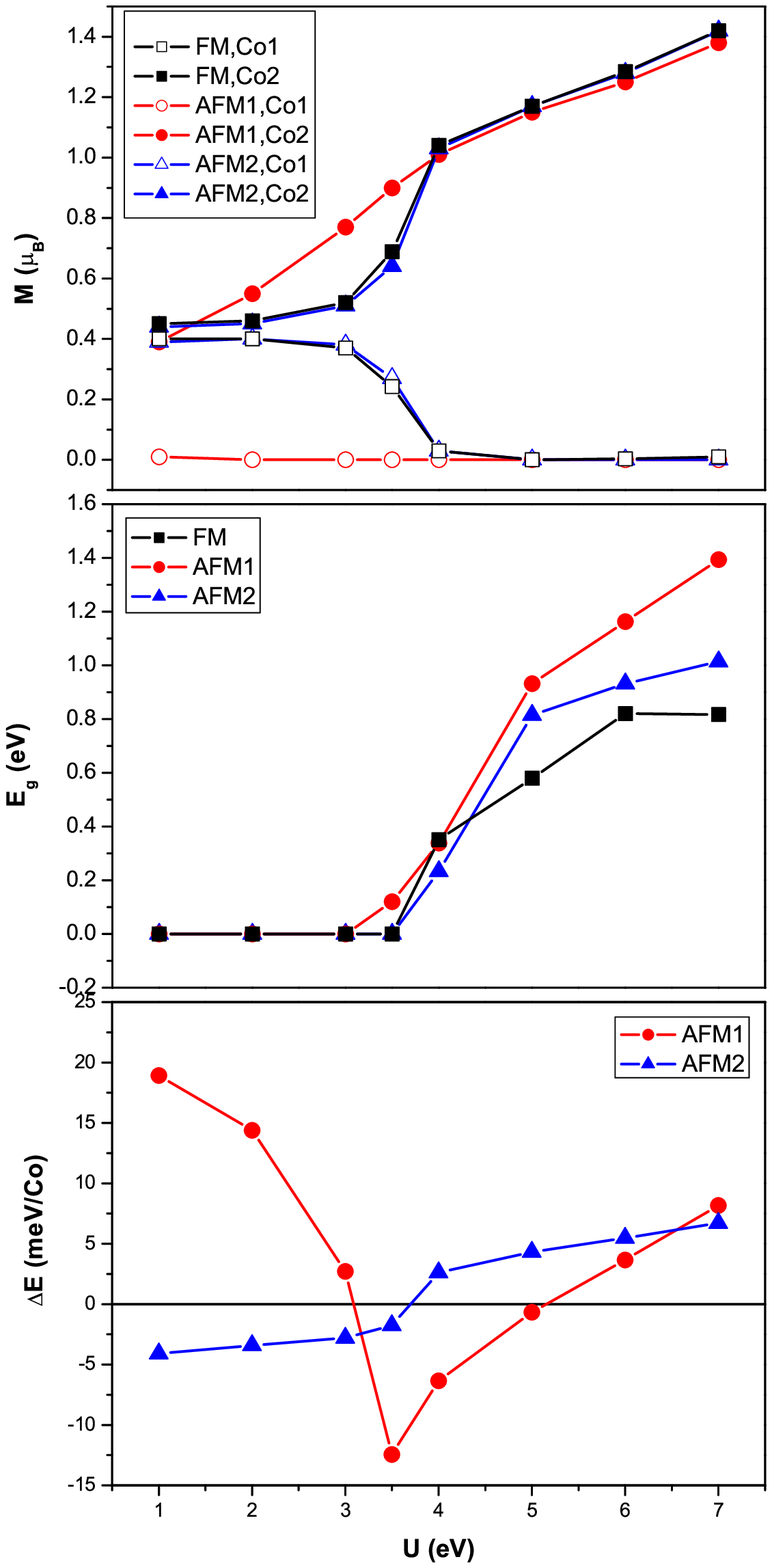}

\end{document}